\documentclass[twocolumn,showpacs,preprintnumbers, amsmath, amssymb,pre]{revtex4-2}
\usepackage{graphicx}% Include figure files
\usepackage{dcolumn}% Align table columns on decimal point
\usepackage{bm}% bold math
\usepackage{amsmath}
\usepackage{color}
\usepackage{float}

\graphicspath{{Figures/}}

\begin{document}
% Lateral contact yields longitudinal cohesion in collective undulatory swimming

\title{Lateral contact yields longitudinal cohesion in active undulatory systems} 
\author{Wei Zhou, Jaquelin Dezha Peralta, Zhuonan Hao, Nick Gravish}
\affiliation{Mechanical and Aerospace Engineering \\  University of California, San Diego}

% Steric synchronization: 

\begin{abstract}

Many animals and robots move using undulatory motion of their bodies. 
When in close proximity undulatory motion can lead to novel collective behaviors such as gait synchronization, spatial reconfiguration, and clustering. 
Here we study the role of contact interactions between model undulatory swimmers: three-link robots in experiment and multi-link robots in simulation. 
The undulatory gait of each swimmer is generated through a time-dependent sinusoidal-like waveform which has a fixed phase offset, $\phi$.
By varying the phase relationship between neighboring swimmers we seek to study how contact forces and spatial configurations are governed by the phase difference between neighboring swimmers.
We find that undulatory actuation in close proximity drives neighboring swimmers into spatial equilibrium configurations  that depend on the actuation phase difference.
We propose a model for spatial equilibrium of nearest neighbor undulatory swimmers which we call the gait compatibility condition, which is the set of spatial and gait configurations in which no collisions occur. 
Robotic experiments with two, three, and four swimmers exhibit good agreement with the compatibility model.
To study the contact forces and configuration equilibrium between undulatory systems we perform simulations.
% The phase variance of a collective governs the allowable packing configurations of undulatory swimmers in agreement with the compatibility model. 
To probe the interaction potential between undulatory swimmers we perturb the each longitudinally from their equilibrium configurations and we measure their steady-state displacement. 
These studies reveal that undulatory swimmers in close proximity exhibit cohesive longitudinal interaction forces that drive the swimmers from incompatible to compatible configurations.
This system of undulatory swimmers provides new insight into active-matter systems which move through body undulation. 
In addition to the importance of velocity and orientation coherence in active-matter swarms, we demonstrate that undulatory phase coherence is also important for generating stable, cohesive group configurations.
\end{abstract}
\maketitle
% \nick{Contact compatibility}

\section{Introduction}

The field of active matter has been inspired by the collective behaviors of biological systems \cite{Vicsek2012-ye}.
The principles of these systems are that individuals move through self-propulsion and that interactions occur through mechanical forces often mediated through hydrodynamic or contact forces \cite{Ramaswamy2010-cz}.
Animal groups across scales from bacteria \cite{Zhang2010-st, Beer2019-kr}, insects \cite{Kelley2013-po, Buhl2006-qf}, fish \cite{Katz2011-bd, Tunstrom2013-sl}, and birds \cite{Castellana2016-py, Cavagna2010-fe} exhibit coordinated movement patterns such as group flocking and swarming. 
In groups of larger animals such as birds and fish the collective movements are generated through visual sensory cues \cite{Ballerini2008-sj,Nagy2010-gb} and hydrodynamic interactions between the individuals \cite{Newbolt2019-gg, Becker2015-xt,Portugal2014-vm,Li2020-nr}.
However, smaller scale systems such as swimming bacteria, sperm, and worms, that often swim in higher group densities may experience repulsive contact forces in addition to fluid interactions \cite{Moore2002-kb, Elgeti2015-mk, Yang2008-hj}. 
The role of contact interactions has been extensively studied in simple models of active matter systems such as self-propelled rods and particles \cite{Shaebani2020-ui, Deseigne2010-di,Aranson2007-en, Kudrolli2008-ru}. 
However, when locomotion is governed by an undulatory motion the interactions between these self-propelled systems may be influenced by phase differences in undulatory gait.
In this work we study how the relationship between spatial configuration and undulatory gait parameters influence the collective behavior of active undulatory systems.

Before introducing active undulatory systems we briefly review the physical phenomena of active-matter and in particular of self-propelled particles. 
A self-propelled particle is an agent that possess an internal energy reservoir which can produce propulsion (see \cite{Bar2020-iv} for an extensive review). 
Groups of these particles can then interact through hydrodynamic, short and long-range potential, or contact forces and display collective behaviors such as flocking, swarming, and incoherent motion. 
Interactions through contact have been extensively studied in these systems and often lead to positional and velocity alignment \cite{Aranson2007-en, Bricard2013-kw,Ginelli2010-vc}. 
In most examinations of the collective physics of self-propelled particles the agents themselves are propelled through constant, time-invariant propulsion.
Steering forces may vary with the environment \cite{Wang2021-ud, Aguilar2018-bu, Campo2010-qt} or the other agents positions (as in the classic Vicsek model \cite{Vicsek1995-lj}), but still typically the propulsion is slowly modulated or constant.
Furthermore, the ``body'' shape of these particles are typically simple spheres, rods, or ellipsoids, that have no articulating components (i.e. are a single rigid body).
This simplification while useful for analysis and simulation is a drastic reduction of the complexity seen in living systems that often locomote through articulated body and appendage motion. 

In this work we define an active undulatory system as consisting of individuals that move through body (or discrete joint) bending that propagates along the length of the body. 
Undulatory locomotion is a common method of movement in biological systems across scales from sperm \cite{Gaffney2011-kh} to snakes \cite{Gray1950-dv, Hu2009-kq,Guo2008-un}.
Undulatory body bending can be three-dimensional with out-of-plane body movement however in this work we consider planar undulatory movement. 
A representative undulatory gait is a simple traveling wave of body bending, $y(x, t) = A \sin(\frac{2 \pi x}{\lambda} - \omega t + \phi)$ that propagates from head to tail. 
The undulatory movement occurs through movement in the lateral direction, $y$, that propagates at wavespeed $\lambda \omega$ and with wavelength $\lambda$ and frequency $\omega$.
However, when considering the undulatory motion of more than one individual, an additional phase parameter $\phi$ becomes necessary to describe the relative motion between the two systems.
When swimmers have identical phase offsets they will be in synchrony, however when phase offsets differ the traveling wave propagation will spatio-temporally differ which might result in forceful interactions between individuals.

\begin{figure}[t]
    \centering
    \includegraphics[width=.77\linewidth]{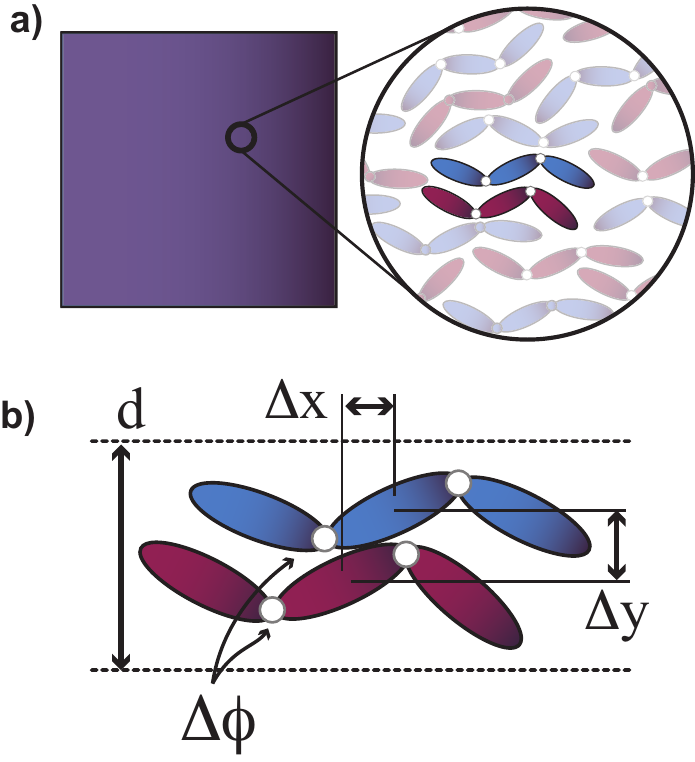}
    \caption{Motivation and overview of gait compatibility among undulatory swimmers. a) Large groups of swimmers experience contact interactions. b) Contact interactions among pairs of undulatory swimmers confined to a lateral distance $d$ require spatial ($\Delta x, \Delta y$) reconfiguration when there is a gait phase difference $\Delta \phi$.}
    \label{fig:fig1}
\end{figure}

The simplest system that can exhibit undulatory, traveling-wave motion is the ``three-link swimmer'' (Fig.~\ref{fig:fig1}).
This system consists of three rigid links separated by two actively controlled joints.
The three-link swimmer was first introduced by Purcell in his study of low Reynolds locomotion \cite{Purcell1977-nk}. 
In the many years since introduction the three-link swimmer has been studied extensively as a model of undulatory locomotion on frictional surfaces \cite{Jing2013-zp, Alben2021-dx}, granular material \cite{Hatton2013-yj}, and within fluids \cite{Tam2007-et,Wiezel2016-ml}.
Undulatory locomotion in a three-link swimmer is generated through oscillatory motion of the two joints, whose angles $[\alpha_1, \alpha_2]$ define a ``shape-space'' of the system \cite{Rieser2019-zh}.
A gait is defined as a closed trajectory through this shape-space over a period of time $T$ such that $\alpha_i(t) = \alpha_i(t + T)$.

Active undulatory systems have been studied in the context of agent-environment interactions such as collision with environmental features. 
Undulatory robots interacting with posts display scattering phenomena that highlight the importance of \textit{active collisions} between active systems and the environment \cite{Rieser2019-pd}.
These authors define active collisions as mechanical contact in which propulsive forces within the robot/animal create persistent contacts with environmental features.
These active collisions yield relationships between the incoming and outgoing trajectory, dependent upon the undulatory phase and collision position.
Similarly microscale swimming bacteria that locomote through reciprocal flagellar movement interact with patterned and flat walls through predictable scattering \cite{Kantsler2013-rp}.
The contact interactions between flagella and the wall redirect the swimmers and the gait-phase at contact governs this scattering behavior.
At a larger scale swimming nematodes (\textit{C. elegans}) make repeated body contact with obstacles when swimming through wet granular material \cite{Juarez2010-tm} and arrays of fixed pillars \cite{Park2008-cv}.
The influence of these obstructions causes the animals to change gait and to generate slower forward velocity.
This previous work highlights how undulatory movement is influenced and affected by interactions with the external environment. 
In particular the importance of gait phase at collision suggests that the phase differences \textit{between} two undulatory swimmers will play an important role in the collective physics of these systems.

This work is inspired by recent observations of collective undulatory swimming in nematodes \cite{Yuan2014-xs}, vinegar worms \cite{Peshkov2021-ji, Quillen2021-fb}, and sperm.
These undulatory swimmers often form clusters of high-density swimmers \cite{Moore2002-kb, Immler2007-oy}, and the close-proximity between individuals can generate forceful interactions through hydrodynamics and contact.
Hydrodynamic interactions between microscale undulatory swimmers have been well studied (see \cite{Elgeti2015-mk} for an extensive review). 
Interactions through a fluid can lead to long and short range forces that drive spatial clustering \cite{Yang2008-hj, Yang2010-ps} and synchronization phenomena \cite{Goldstein2009-ou, Brumley2016-jb, Brumley2014-cg, Guo2018-tw,Kotar2010-rf}. 
When organisms increase in size the role of hydrodynamic interactions is diminished, yet individuals may still interact through contact.
Recent experiments with vinegar worms \cite{Peshkov2021-ji, Quillen2021-fb} and nematodes \cite{Yuan2014-xs} demonstrate that contact interactions can generate synchronization of the undulatory gait.
More broadly, contact interactions between undulatory systems can generate coherent and incoherent movement dependent on density, gait, and actuation parameters \cite{Chelakkot2020-fq, Reigh2012-km, Ozkan-Aydin2021-br}.

\begin{figure*}[t]
    \centering
    \includegraphics[width=1\linewidth]{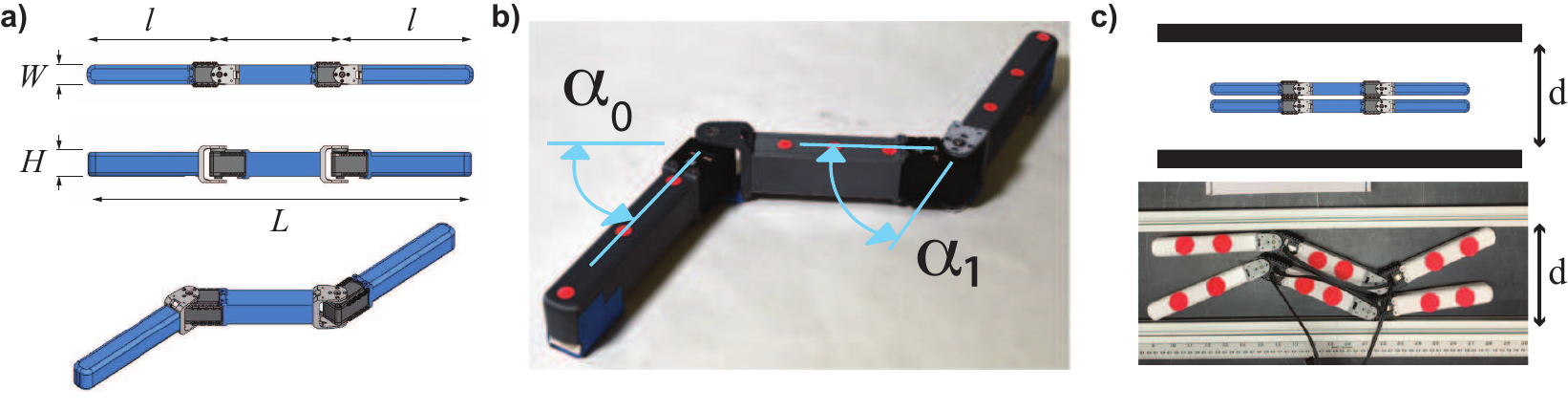}
    \caption{Overview of three-link robots. a) Geometry of the three-link system. Links are length, $l = 18$~cm, height, $h = 5$~cm, and width, $w = 1.5$~cm. b) The joint angles, $\alpha_0$ and $\alpha_1$ are controlled through position-commanded servos. c) We studied groups of robots in a narrow channel of variable width $d$.}
    \label{fig:methods}
\end{figure*} 

This paper is organized through a series of robophysical and simulation experiments to probe the spatial dynamics of active undulatory systems.
In Sec.~\ref{sec:methods} we provide details for robot design and experimental parameters. 
The first observations are made with pairs of undulatory robots in~Sec.\ref{sec:two_robots} that demonstrate longitudinal re-positioning when gait phases differ. 
From these experiments we develop a model in Sec.~\ref{sec:gait_compatibility} of spatial configurations that depend on gait which we call \textit{gait compatibility}.
The gait compatibility model is based on an assumption of a smooth, sinusoidal body shape that differs from the three-link robots studied in experiment and simulation and thus we next study the role of body shape on compatibility (Sec.~\ref{sec:more_links}).
We next study groups of three and four swimmers to observe spatial dynamics in these collectives (Sec.~\ref{sec:four_robots}).
The group experiments and the compatibility model suggest a packing density limitation with groups of swimmers at different phases which is investigated in Sec.~\ref{sec:packing}.
Lastly, we observe the basins of attraction of compatible configurations (Sec.~\ref{sec:basins}) and we measure the ``potential energy'' of these stable configurations in simulation (Sec.~\ref{sec:potential}).
This combination of experiments and simulation reveal that undulatory phase differences have a significant influence on spatial configurations within undulatory active systems.

\section{Methods}
\label{sec:methods}
We performed experiments and simulation of undulatory robots that have three rigid links coupled through two rotational joints. In both experiment and simulation the rotational joints are commanded by angular position trajectories. We studied the behaviors of two and up to ten robots through experiment and simulation (Fig.~\ref{fig:methods}). 

\subsection{Robotics experiments}

The experimental set-up was designed to observe the collision of two three-link robots interacting within a confined area. 
Each robot had a 3D printed body connected with two Dynamixel 12-A servo motors (Fig.~\ref{fig:methods}a) with a total length, $L = 50$~cm.
All servo motors were programmed to oscillate with a sine function of constant amplitude and constant frequency (0.5 Hz). 
To produce an undulatory motion, we generated a traveling wave along the length of the body with angular position of the $i^{\text{th}}$ joint on the $j^{\text{th}}$ robot as
\begin{align}
    \alpha_i^j = \beta \sin\left(\frac{2\pi L i }{\lambda N} - 2\pi f t + \phi_j \right)
    \label{eqn:waveform}
\end{align}
where $i = [0, 1]$ denotes the joint number and $N = 3$ for the three links of the robot (Fig.~\ref{fig:methods}b). 
The ratio $\frac{\lambda}{L}$ is the normalized wavelength of the actuation wave.
In our experiments we varied $\frac{\lambda}{L} = [0, \frac{2}{3}, 1, \frac{4}{3}]$.
The angular amplitude, $A$, was held constant at $\beta = 45^\circ$ in experiment and varied between $\beta = 30-60^\circ$ in simulation. 
The phase offset $\phi_j$ is constant for each robot, but could differ between robots, and represents the overall actuation phase of the robot. 
Thus the phase difference between robots is represented by $\Delta \phi = \phi_a - \phi_b$.

The goal of this paper is to study the spatial dynamics of undulatory swimmer groups as they swim in the same direction (Fig.~\ref{fig:fig1}a). 
We emulated the effect of being within a group by confining the robots to a narrow channel so that they are forced to interact with each other. 
When the confining wall is not in place the robots will push each other away until they no longer contact and interact.
The confined environment was created using a fixed channel measuring (slightly more than) one meter long and 13 cm wide (Fig.~\ref{fig:methods}c). 
The robots rested on a smooth frictional surface and were confined laterally by two rigid walls whose separation distance, $d$, was varied depending on the number of robots in the experiment. 
When robot pairs or groups are in the channel they are able to move laterally ($\Delta  y$) and longitudinally ($\Delta x$) with respect to each other through contact. 
Each experiment consisted of placing the robots laterally in contact ($\Delta y = 1.5$~cm) and at the same longitudinal position ($\Delta x = 0$). 
Video recording of the robot movements was captured from an overhead view.
A meter stick was aligned along the channel’s length to measure the robot locations measured at each robot’s center. 
At the start of each experiment, the two robots were set to their elongated shape, with joint angles $\alpha_i^j = 0$. 
 
\begin{figure*}[t]
    \centering
    \includegraphics[width=1\linewidth]{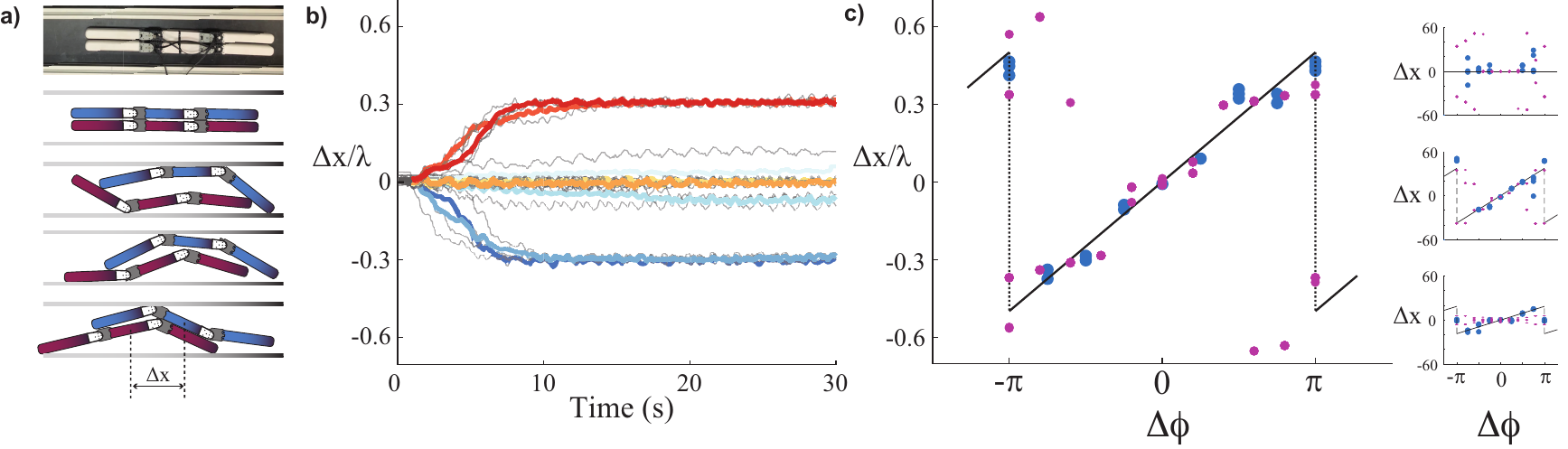}
    \caption{Gait compatibility in undulatory swimmer pairs. a) Image of two undulatory robot swimmers in experiment and illustration of longitudinal motion from contact interactions. Illustrations are traced from experiment images. b) Longitudinal separation distance versus time for robot pairs in simulation. c) Steady-state longitudinal separation versus phase difference between robot pairs with $\frac{\lambda}{L} = 1$. Large blue dots are experiment results; smaller purple dots are simulation results. Lines are compatibility predictions from equation~\ref{eqn:compatibility}. The right column shows the non-normalized displacements ($\Delta x$) with from experiments and simulation with wavelengths $\lambda/L = [0, \frac{4}{3}, \frac{2}{3}]$. The lines are compatibility prediction.}
    \label{fig:two_robot_experiment}
\end{figure*} 

Through video tracking we measured the center position of all robots in experiment. 
We compute the center-to-center spacing between nearest neighbors to determine the lateral and longitudinal spatial shifts that occur during undulatory movement. 
The motors were commanded to generate an undulatory gait for 15 oscillations (30 seconds). 
In robot pairs this process was repeated for the variable phase shift in the two and three robot experiments, $\phi$, at values [-1, $-\frac{3}{4}$, -$\frac{1}{2}$, -$\frac{1}{4}$, 0, $\frac{1}{4}$, $\frac{1}{2}$, $\frac{3}{4}$, 1]$\pi$. 
For the four robot tests the phase shifts were selected at random.
We additionally performed this measurement for varying wavelengths.

The robots did not have wheels or any other frictional anisotropy and thus they do not ``locomote'' but rather undergo continuous traveling wave oscillation while approximately remaining in the same spatial location. 
While the robots do not swim in experiment, our focus in this manuscript is the influence of contact interactions on the relative spatial positioning between undulatory swimmers. 
In simulations we implemented viscous drag forces on the swimmers to make them swim forward and interact with each other through contact and we found no qualitative difference in spatial dynamics between the locomoting, and non locomoting simulations. 
Thus, despite the robots not swimming through a fluid we believe their contact interactions are commensurate with those of the swimming systems that inspired this work.

\subsection{Time-dependent joint synchronization simulation}

We performed simulations of undulatory robot groups to compare with experiment and to extend analysis of this system beyond what is experimentally feasible.
We used two simulation environments to study this system: 1) the Matlab Simscape Multibody physics environment with contact forces, and 2) the Project Chrono multibody physics simulation environment \cite{Chrono2016}. 
The robots interacted with normal forces only, friction between the robots was ignored. 
To model the friction between the robot and ground we implemented viscous damping on the $x,~y$ motion of each link of the robot. 
Viscous damping simplified the computation while maintaining the same overall behavior as in experiment: the movement of robots on the table is overdamped and robots rapidly come to rest when pushed away from each other. 
We modeled the position controlled servos in simulation through torque actuation of the joints under a PID position control with torque saturation (to model the maximum torque capabilities of the motors).
The motors were commanded to follow the trajectories defined in Equation~\ref{eqn:waveform}.
We primarily studied groups of 3-link robots consistent with experiment.
However, we did perform simulations with 5-link, 7-link, and 9-link robots to understand how link number influenced spatial dynamics. 

\section{Results and discussion}

\subsection{Spatial reconfiguration between robot pairs}
\label{sec:two_robots}
We begin our investigation by studying pairs of robots in experiment and simulation. We set the channel width to $d = 13$~cm to constrain the robots laterally and we perform experiments with a phase difference between the two robots of $\Delta \phi = \phi_1 - \phi_2$ over the range of $\Delta \phi \in [-\pi, \pi]$. 
We observe that robots with non-zero $\Delta \phi$ experience a longitudinal displacement, $\Delta x$, driven by the contact interactions (Fig.~\ref{fig:two_robot_experiment}a,b). 
The longitudinal separation, $\Delta x$, versus time from identical simulations indicates that there is an initial rapid rate of separation which eventually slows until reaching what appears to be the steady-state value for longitudinal spacing (Fig.~\ref{fig:two_robot_experiment}b). 

In experiments where the phases of the two robots differed the steady-state longitudinal separation is approximately linear with the phase difference (Fig.~\ref{fig:two_robot_experiment}c).
The slope of the $\Delta x$ versus $\Delta \phi$ relationship was dependent on the number of traveling waves along the body $\xi = [0, \frac{3}{4}, 1, \frac{3}{2}]$ (Fig.~\ref{fig:two_robot_experiment}c, right panel). 
In simulation we conducted identical experiments which reproduced the longitudinal separation (Fig.~\ref{fig:two_robot_experiment}c).
Simulation also allowed us to measure the contact forces exerted between robots during experiment and we found that the equilibrium conditions coincided with a reduction in the contact forces. 
When robots were in the steady-state phase the contact forces were small or non-existent for some phase-displacement configurations. This observation indicates two important points: 1) as expected contact interactions drive the spatial dynamics in this system, and 2) the system evolves to a state which minimizes the contact interactions among these active undulatory robots.

\subsection{A gait compatibility model for undulatory collectives}
\label{sec:gait_compatibility}

We hypothesize that undualatory swimmers actuated through a sinusoidal traveling wave adjust their spatial positioning to minimize contact interactions. 
We now derive a geometric relationship between phase and spatial configuration based on the assumption of minimizing contact. 
We assume that the undulating motion of the 3-link robot is represented by a sinusoidal traveling wave of amplitude, $A$, and wave length, $\lambda$ and that there is no lateral separation between the robots ($\Delta y = 0$; we will relax this assumption later). 
In the continuum limit the lateral position of each robot is thus described by, 
\begin{align}
y_1(x,t) &= A \sin\left(2\pi\frac{x}{\lambda} - \omega t \right) \nonumber \\
y_2(x,t) &= A \sin\left(2\pi\frac{(x - \Delta x)}{\lambda} - \omega t + \Delta\phi \right) \nonumber
\end{align}

\begin{figure}[t]
    \centering
    \includegraphics[width=1\linewidth]{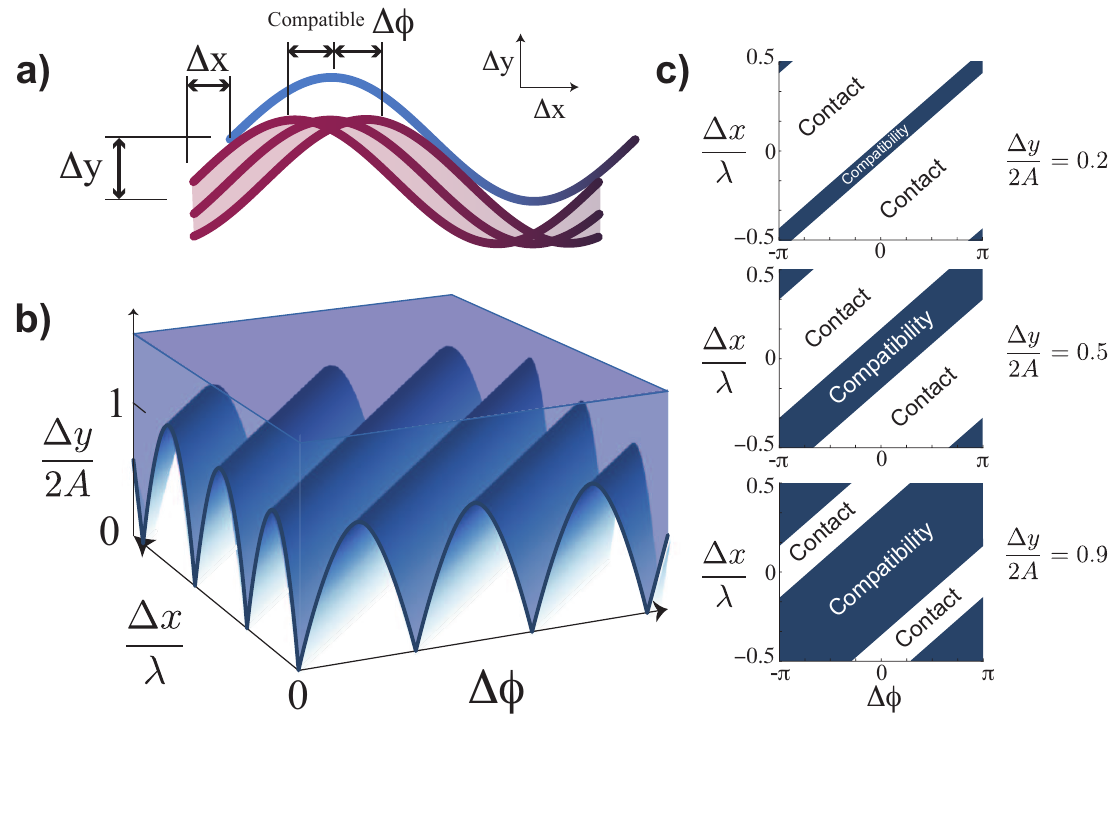}
    \caption{Contact compatibility criteria. a) An overview of the phase-range ($\Delta \phi$) for compatible sinusoidal curves that are separated by a $\Delta x$ and $\Delta y$ displacement. b) The three-dimensional representation of the compatibility criteria represented by equation~\ref{eqn:fullcompatibility}. The shaded region above the solid blue curves are allowable configurations of gait compatibility. As the lateral separation distance increases (vertical axis, $\frac{\Delta y}{2A}$) the range of compatible phases increases. c) We show cross sections of the compatibility condition at three different lateral separations: $\frac{\Delta y}{2A} = [0.2, 0.5, 0.9]$ from top to bottom. When $\frac{\Delta y}{2A} > 1$ any combination of $\Delta \phi$ and $\Delta x$ will be in compatibility.}
    \label{fig:compatibility}
\end{figure}

We propose that the equilibrium configuration of undulatory swimmers occurs when the two sinusoidal curves make tangential contact (i.e. they are just close enough to touch but do not intersect) which is shown in Fig.~\ref{fig:compatibility}a). 
For the case of no lateral separation this imposes the single constraint, $y_1(x,t) = y_2(x,t)$, which can be satisfied by a relative longitudinal displacement between the two robots by an amount
\begin{align}
    \label{eqn:compatibility}
    \Delta x = \frac{\lambda}{2 \pi} \Delta \phi
\end{align}
We call this condition the compatibility condition for undulatory swimmers, inspired by recent experiments on swimming worms which introduced the term gait compatibility \cite{Yuan2014-xs}. 

We plot the gait compatibility prediction along with the experimental measurements in Figure~\ref{fig:two_robot_experiment}~c,d for the robot pairs. 
We find good agreement with the model prediction across the four actuation wavelengths explored in experiment. 
Similarly in simulation we find good agreement between the model and observation, with some systematic variation that is a result of the sinusoidal assumption for the three-link robot. 
The gait compatibility model suggests that oscillatory robots with traveling wave actuation can swim in close proximity by adjustments to their longitudinal position according to their phase difference. 
Critically, when the undulatory body wave was stationary ($\frac{\lambda}{L} = 0$) the robots and simulation displayed large variance in $\Delta x$. 
This suggests the importance of traveling wave actuation which acts to couple the lateral contact with longitudinal reconfiguration. 
In effect the traveling wave actuation can force neighboring swimmers along the longitudinal axis and drive them into compatible states. 

In the deriving Equation~\ref{eqn:compatibility} we did not consider the influence of a lateral separation distance, $\Delta y$, on the allowable phase and longitudinal offsets in which compatibility is achieved. 
However, in larger groups contact interactions may lead to density fluctuations \cite{Narayan2007-yw}.
These density fluctuations may increase the range of compatible $\Delta \phi$.
Here we now derive the full compatibility relationship that governs the allowable lateral, longitudinal, and phase offsets for two compatible sinusoidal gaits.

\begin{figure*}[t]
    \centering
    \includegraphics[width=1\linewidth]{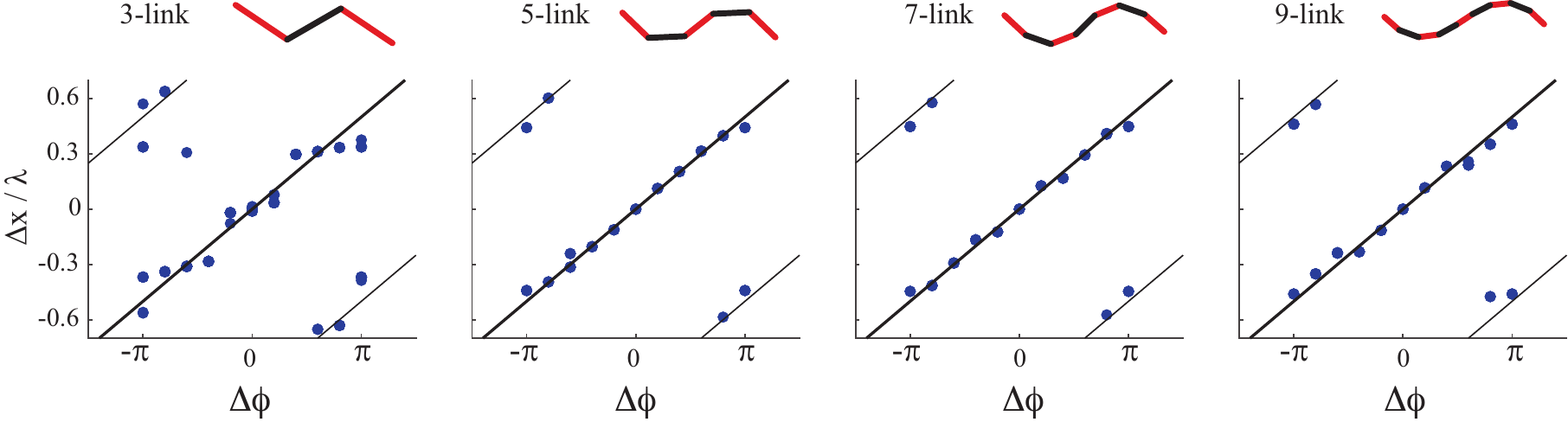}
    \caption{Gait compatibility simulations of two robots with 3, 5, 7, or 9 links from left to right. As the link number increases the longitudinal separation versus phase difference (circles) show good agreement with the compatibility prediction (lines).}
    \label{fig:more_links}
\end{figure*}

We again assume two swimmers oscillate as spatial sinusoidal waves and we now include the longitudinal ($\Delta x$), phase $\Delta \phi$, and lateral ($\Delta y$) offsets
\begin{align}
    \label{eqn:sin1}
    y_1(x,t) &= A \sin\left(2\pi\frac{x}{\lambda} - \omega t \right) \\
    y_2(x,t) &= A \sin\left(2\pi\frac{(x - \Delta x)}{\lambda} - \omega t + \Delta\phi \right) + \Delta y 
    \label{eqn:sin2}
\end{align}
We assume that $\Delta y > 0$ and thus swimmers that are in compatibility satisfy the equation $y_1(x, t) \leq y_2(x,t)$ (Fig.~\ref{fig:compatibility}a). 
However, the boundaries of the compatible states occur when two sinusoidal curves make tangential contact, which imposes the following two constraints, $y_1(x^*,t) = y_2(x^*,t)$, and $y_1'(x^*,t) = y_2'(x^*,t)$ where prime denotes derivative with respect to $x$ and $x^*$ is the contact location. 
Solving this equation for the allowable phase offsets between neighboring swimmers yields the following inequality relationship
\begin{align}
    |\Delta \phi - \frac{2\pi}{\lambda}\Delta x| \leq ~2 \left|\arcsin{\left(\frac{\Delta y}{2A}\right)}\right|
    \label{eqn:fullcompatibility}
\end{align} 
When $\Delta y = 0$ this yields the previous equality in Eqn.~\ref{eqn:compatibility} between $\Delta x$ and $\Delta \phi$.
However, as $\Delta y$ increases there is a growing range of allowable phase offsets in which two undulatory swimmers can be in gait compatibility and not make contact.

We show samples of the compatibility condition for varying ranges of lateral separation in Fig.~\ref{fig:compatibility}b,c which highlights the growing region of compatible phase and longitudinal separation as lateral distance increases. 
The gait compatibility described here is somewhat similar to the hydrodynamic synchronization of infinite two-dimensional undulating sheets studied by Elfring and Lauga \cite{Elfring2009-ze}. 
In that work, the two sheets could freely displace longitudinally through fluid force interactions and the authors demonstrated that the sheets always converged to a relative in-phase or anti-phase configuration depending on waveform.
However, there is a critical difference between this work and the case of systems that interact through hydrodynamic forces.
When two swimmers are within gait compatibility, they do not contact each other and thus are entirely decoupled. 
Small perturbations to their position or phase, as long they are not pushed out of compatibility, will persist and thus the compatibility state is a neutrally stable configuration.  
The interaction through contact means that there is a discontinuity at the boundary between states where the swimmers can interact, and states where they cannot interact.

\subsection{Increasing link number yields better agreement with compatibility model}
\label{sec:more_links}

In both experiment and simulation we noticed that the equilibrium $\Delta x$ formed discrete clusters along the compatibility prediction line while the prediction from a sinusoidal model is a linear phase-displacement relationship. 
We hypothesized that this model error was the result of the poor approximation of a sinusoidal shape by the 3-link system. 
We studied 5, 7, and 9-link robot pairs in simulation and we found that increasing the linkage number produced an increasingly linear compatibility relationship with increasing link number. 
The root mean square (RMS) error of the simulation compatibility separation and the prediction (Eqn.~\ref{eqn:compatibility}) decreased with increasing link number (RMS error from compatibility condition = [0.056, 0.042, 0.025, 0.039] for the [3,5,7,9] link robots respectively). 
The decreasing error with increasing link number is a result of the discretized body-shape in the three-link robot experiments. 
The compatibility model assumes a perfectly sinusoidal body shape however with only three-links the body undulation is not quite sinusoidal.
However, as we add more links this assumption becomes better so does the gait compatibility model.

\subsection{Experiments with three and four robot pairs}
\label{sec:four_robots}
\begin{figure}[t]
    \centering
    \includegraphics[width=1\linewidth]{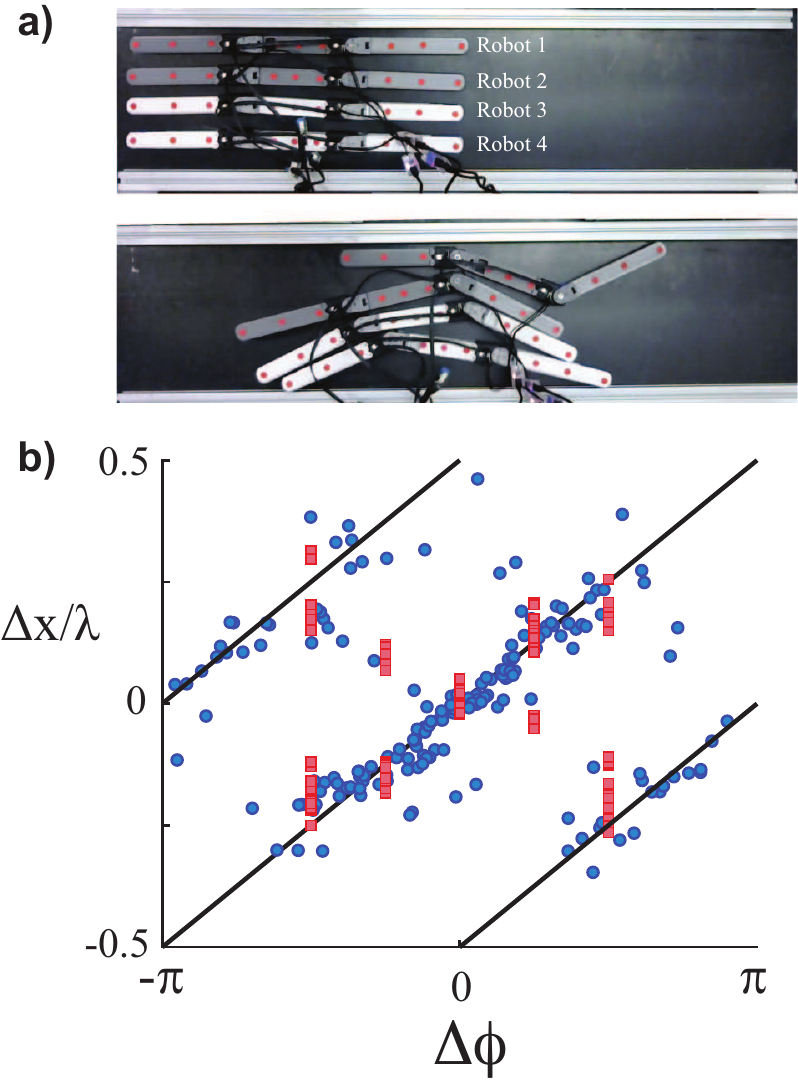}
    \caption{Gait compatibility in larger robot groups. a) Image of a four robot experiment at the beginning (top) and in the middle of the experiment (bottom). b) Steady-state longitudinal separation versus phase difference groups experiments with three (red squares) and four robots (blue circles). Lines are compatibility predictions from equation~\ref{eqn:compatibility}.}
    \label{fig:robotgroups}
\end{figure}

To examine how larger groups of undulatory swimmers spatially organize we performed experiments with groups of three and four robots (Fig.~\ref{fig:robotgroups}a). 
We widened the channel to $d = 19~\&~22$~cm for three and four robot experiments and set $\frac{\lambda}{L} = 1$. 
The robots are initialized with $\Delta x = 0$ and all joint angles set to zero. 
We begin undulatory actuation for the robots and we monitor the lateral and longitudinal displacement from an overhead camera (Fig.~\ref{fig:robotgroups}a).
We observed the same overall behavior in robot groups as in robot pairs: phase differences between neighboring robots resulted in longitudinal repositioning until the group reached an overall steady-state spatial configuration. 
Examining the nearest neighbor $\Delta x_{i,i+1}$ versus $\Delta \phi_{i,i+1}$ we observe good agreement with the compatibility predictions (Fig.~\ref{fig:robotgroups}b). 
The variation in the longitudinal position was larger in the group experiments compared with the robot pair experiments suggesting potential collective effects present in the three and four robot experiments that were not captured in the pair experiments. 
The principle influence of this variance from the compatibility prediction is the larger lateral spacing afforded to the larger robot groups. 
As the swimmers push each other they may arrange into higher density configurations leaving lateral space for the others, which thus increases the range of compatible phases allowable (Eqn.~\ref{eqn:fullcompatibility}). 
We will explore how density influences phase variance in the next section.

\subsection{Gait compatibility influences spatial packing}
\label{sec:packing}
\begin{figure}[t]
    \centering
    \includegraphics[width=1\linewidth]{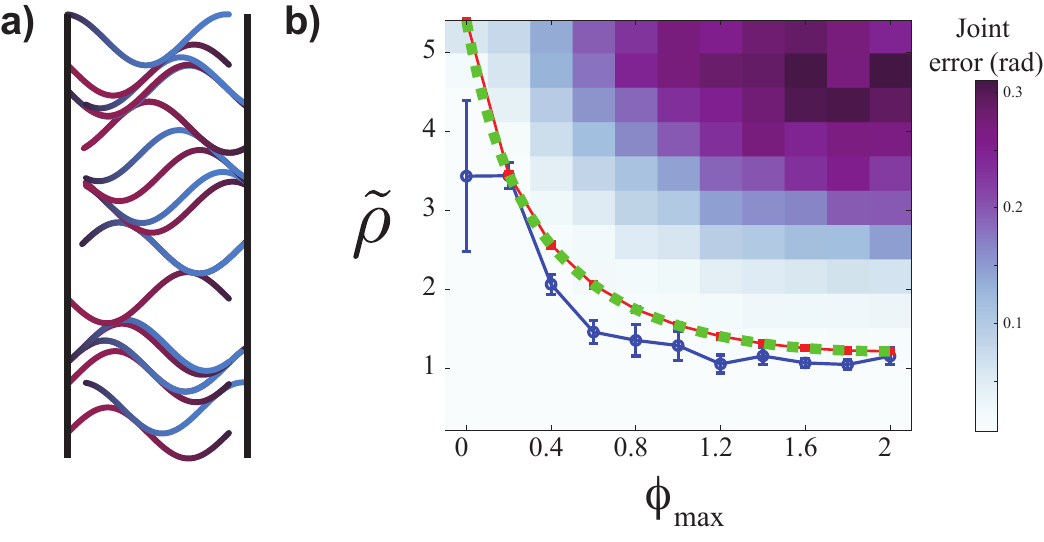}
    \caption{Lateral density is influenced by phase variance in undulatory groups. a) A representation of a group of undulatory swimmers separated laterally (the vertical direction) with phase variation. b) A heatmap of the cumulative joint error induced through collisions within a robot group for varying normalized lateral density, $\tilde{\rho}$ (y-axis) and phase variance (x-axis). Blue circles indicate the measured threshold below which contact typically does not occur in the group from simulations. The red curve is a Monte-Carlo estimate based off of the compatibility equation. The dashed green curve is the calculation from the math model Eqn.~\ref{eqn:rho_tilde}.} 
    \label{fig:spatial_packing}
\end{figure}

Contact interactions among the collective undulatory swimmers drive them into compatible configurations. 
However, the range of available compatible configurations increases as the lateral spacing increases and thus we expect that the group density will influence the allowable phase and spatial variance in groups. 
To address this question we studied the packing configurations of large groups of three-link swimmers in simulation to determine the relationship between group phase variance and packing density. 
We used a short channel to confine the swimmers longitudinally by a distance of 1.1$L$ constraining the ability to longitudinally reconfigure. 
We initialized 50 robots within a channel of fixed lateral width (Fig.~\ref{fig:spatial_packing}a) and we measured the spatial positioning, and the deflection of the rotational joints from their commanded trajectory (joint error). 
The swimmers were all oriented along the direction of the channel length so that they only interacted through lateral collisions and their orientations were approximately the same. 
We varied the lateral confinement distance and the range of gait phases to observe how spatial and phase variance influences the packing and contact interactions between groups of undulatory swimmers.

We characterized contact interactions between robots by monitoring the instantaneous joint error from their setpoint angular position.
The joint error is linearly proportional to the joint torque in the joint control system, and thus this is a metric of contact interactions between robots.
When the robots are not in contact the joint error is approximately zero however when the lateral density is increased the robots begin contacting each other and causing the joints to deviate from their assigned motion.
We examine the influence of phase-range and lateral density on the packing behavior of the robots (Fig.~\ref{fig:spatial_packing}b).
For a fixed density, increasing the phase-range resulted in an increase in the overall joint error of the robots in the group indicating collisions and non-compatible space-phase relationships.
Similarly, for a fixed phase-range increasing the density caused an increase in joint error. 

We characterized the compatibility threshold in simulations by determining for each phase-range and density combination whether the mean joint-error was above that of an individual robot (threshold of 0.01~rad.). 
The compatibility threshold from simulations is shown in blue circles in Figure~\ref{fig:spatial_packing}b, where error bars are the result of 5 replicate simulations.
We similarly computed the compatibility threshold for sinusoidal swimmers using a Monte-Carlo approach where we computed the required density to maintain compatibility for swimmers with random phase range between 0 and $\phi_{max}$ (red curve, Fig.~\ref{fig:spatial_packing}b).
Lastly, we can exactly compute the compatibility packing threshold for a group of swimmers with random, uniform phase distribution in the range $\phi_i = [0, \phi_{max}]$ using Equation~\ref{eqn:fullcompatibility}.
We set $\Delta x = 0$ and rearrange Equation~\ref{eqn:fullcompatibility} to the following 
\begin{align}
    \frac{\Delta y}{2 A} = \sin\left(\frac{\Delta \phi}{2}\right)
    \label{eqn:sep}
\end{align}
For a swimmer group with phases drawn at random from the uniform distribution $\phi_i = [0, \phi_{max}]$ the expected robot separation in the y direction can be derived from Equation~\ref{eqn:sep}
\begin{align}
    \sum_{i=1}^{N} \frac{|\Delta y_i|}{2AN} &=  \frac{1}{N} \sum_{i=1}^{N} \left|\sin\left(\frac{ \phi_{max}}{2N}i\right)\right| \cdot f_{pdf}\left(\frac{\phi_{max}}{N}i \right)
\end{align}
using the probability distribution of the phases, $f_{pdf}$.
We take expected value to the continuum limit ($N \rightarrow \infty$ as
\begin{align}
     \frac{\tilde{y}}{2A} &= \frac{1}{\phi_{max}} \int_{0}^{\phi_{max}} \left|\sin\left(\frac{x}{2}
    \right)\right| \, f_{pdf}(x) \,dx \\
     &= \frac{1}{\phi_{max}} \int_{0}^{\phi_{max}} \left|\sin\left(\frac{x}{2}
    \right)\right| \, (2 - \frac{2}{\phi_{max}}x) \,dx \\
     &= \frac{4}{\phi_{max}} - \frac{8 \sin\left(\frac{\phi_{max}}{2}\right)}{\phi_{max}^2} \\
     &= \frac{4}{\phi_{max}^2}\left(\phi_{max} - 2 \sin \left(
     \frac{\phi_{max}}{2} \right)\right)
\end{align}

We now calculate the expected lateral distance ($Y$) required for a group of $N$ robots, considering the robot body width $w$
\begin{align}
    Y &= N\tilde{y} + N w \\
      &= \frac{AN 8}{\phi_{max}^2}\left(\phi_{max} - 2 \sin \left(
     \frac{\phi_{max}}{2} \right)\right) + Nw
\end{align}
Thus the expected lateral density with all swimmers in compatibility is $\rho = N/Y$ and the density normalized by the peak-to-peak oscillatory amplitude is given by $\tilde{\rho} = 2AN/Y$ which yields 
\begin{align}
    \tilde{\rho} &= \frac{1}{\frac{4}{\phi_{max}} - \frac{8 \sin(\frac{\phi_{max}}{2})}{\phi_{max}^2} + \frac{w}{2A}} \label{eqn:rho_tilde}
    % &= \frac{2 A \phi_{max}^2}{8 A \phi_{max} + \phi_{max}^2 w - 16 A \sin\left(\frac{\phi_{max}}{2}\right)}
\end{align}
We see in Figure~\ref{fig:spatial_packing}b that the expected value calculation agrees extremely well with the Monte-Carlo simulation.
Furthermore, we can examine the extremes of the density variation and their effect on packing density. 
When $\phi_{max} = 0$ the non-normalized density becomes $\rho = \frac{1}{w}$ which corresponds to the maximum packing density of the robots in contact with each other.
Overall we find that as the phase variation within a swimming group increases the required lateral density within the group must increase or else swimmers will collide with each other. 
This relationship provides compelling motivation for animal and engineered swarms of swimming agents to synchronize their gaits to achieve higher density groups.

\label{sec:basins}
\begin{figure*}[t]
    \centering
    \includegraphics[width=1\linewidth]{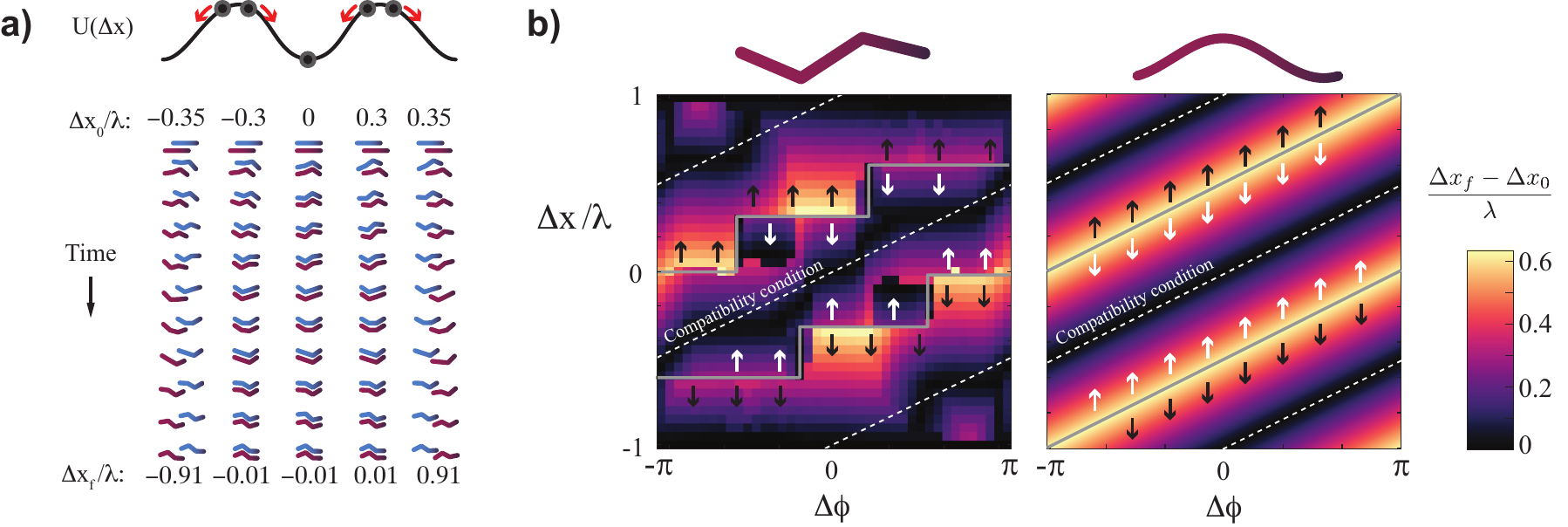}
    \caption{Compatible configurations minimize the contact between swimmers. a) We envision that longitudinal dynamics are governed by a potential energy landscape dependent on the phase difference between swimmers. Swimmers initialized with $\Delta \phi = 0$ and different longitudinal positions ($\Delta x_0 / \lambda$) evolve to one of three compatible configurations dependent on initial position. All initial positions $|\Delta x_0 / \lambda | \leq 0.3$ evolve to $\Delta x_f / \lambda = 0$ b) Heatmap represents the distance traveled from initial condition to compatibility, $|\Delta x_f - \Delta x_0|$ for three-link swimmers (left) and sinusoidal swimmers (right).} 
    \label{fig:basins}
\end{figure*}

\begin{figure}[t]
    \centering
    \includegraphics[width=0.8\linewidth]{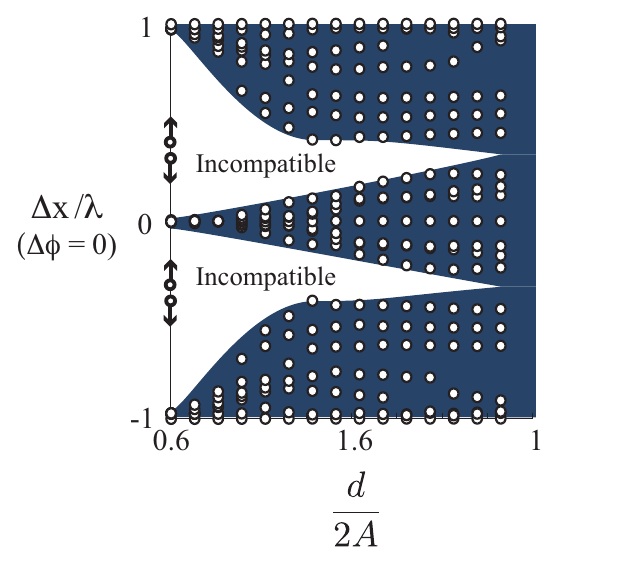}
    \caption{Steady state longitudinal separation between two swimmers with $\Delta\phi = 0$ as a function of lateral confinement. } 
    \label{fig:spacing_basins}
\end{figure}
\subsection{Compatible configurations have a broad basin of attraction}
Swimmers in non-compatible configurations are rapidly pushed into compatibility through contact interactions (Fig.~\ref{fig:two_robot_experiment}b). 
In the idealized situation in which swimmers are represented by sinusoidal body position the compatible configuration is an equilibrium condition where contact no longer occurs. 
In this section we study how the initial longitudinal separation, $\Delta x$, and phase detuning, $\Delta \phi$, influence the final state reached by the pair of undulatory swimmers. 
We study this for the case of close proximity with wall separation $d = 13$~cm where contact interactions are reinforced by the close proximity (i.e. swimmers cannot push each other away laterally).

In a first example we study the spatial evolution of five different initial conditions of longitudinal separation, $\Delta x_0$ (Fig.~\ref{fig:basins}a).
The phases are the same for these swimmers ($\Delta \phi = 0$) and so the compatible configuration is $\Delta x = 0$. 
Initial separation distances that are far away from the compatible separation ($|\Delta x_0| > 0.3$) are pushed away from the $\Delta x = 0$ compatible configuration as the swimmers repel each other along the longitudinal axis.
However, when the initial separation distances are closer (approximately $|\Delta x_0| \leq 0.3$) the swimmers experience an attractive interaction force in the longitudinal direction and ultimately end in the compatible state for there phase difference. 

We study the evolution of longitudinal separation across the full range of relevant initial separation and phase differences.
From each initial condition we compute the total longitudinal position change, $\Delta x_f - \Delta x_0$ and plot the heatmap of this value (Fig.~\ref{fig:basins}b).
When the position change is positive the swimmers experience a net longitudinal repulsion and move away from each other, while negative values indicate that the swimmers experience a net attractive interaction and move towards the compatible configuration. 
It is interesting to note that while the only interactions between the swimmers are through repulsive contact forces, the confinement and the traveling wave shape change results in regions of longitudinal attraction between the swimmers.  
This attractive potential will be further studied in the next section. 

From the position change map we can clearly see why the three-link swimmer experiment exhibited the clustering along the $\Delta x$ axis (Fig.~\ref{fig:two_robot_experiment}b, and Fig.~\ref{fig:more_links}a).
The minimum of the interaction ``potential'' does not follow the compatibility prediction from the sinusoidal calculation, and instead follows a stair-stepped shape along the $\Delta x$ vs. $\Delta \phi$ parameter space (Fig.~\ref{fig:basins}b). 
To determine how different the three-link swimmer position change map is from the sinusoidal prediction we compute the longitudinal position change map for two sinusoidal swimmers. 
We assume a lateral separation $\Delta y \neq 0$, and that the amplitude of two infinitely long sinusoidal curves grows in time from $A = 0$ at initiation to $A_f$ following Equations~\ref{eqn:sin1}~\&\ref{eqn:sin2}. 
As the curves grow in amplitude they will eventually make tangential contact at the point, $x^*(t) = \frac{\Delta}{2\pi} - \frac{\Delta \phi \lambda}{4 \pi} - \omega t$, which is given by the same conditions used to determine compatibility (see Section~\ref{sec:gait_compatibility} and Equation~\ref{eqn:fullcompatibility}).
The normal vector between the tangent point of the two curves will determine the direction the curves will be pushed as the amplitude continues to grow. 
The distance the curves must displace to reach compatibility is minimum distance from the contact location to the large amplitude compatibility condition given by Equation~\ref{eqn:compatibility}.
In Figure~\ref{fig:basins}b we show the displacement required to reach compatibility for the sinusoidal system. 
The basins of attraction for the different compatibility lines are lines of the same slope.
\begin{figure*}[t]
    \centering
    \includegraphics[width=1\linewidth]{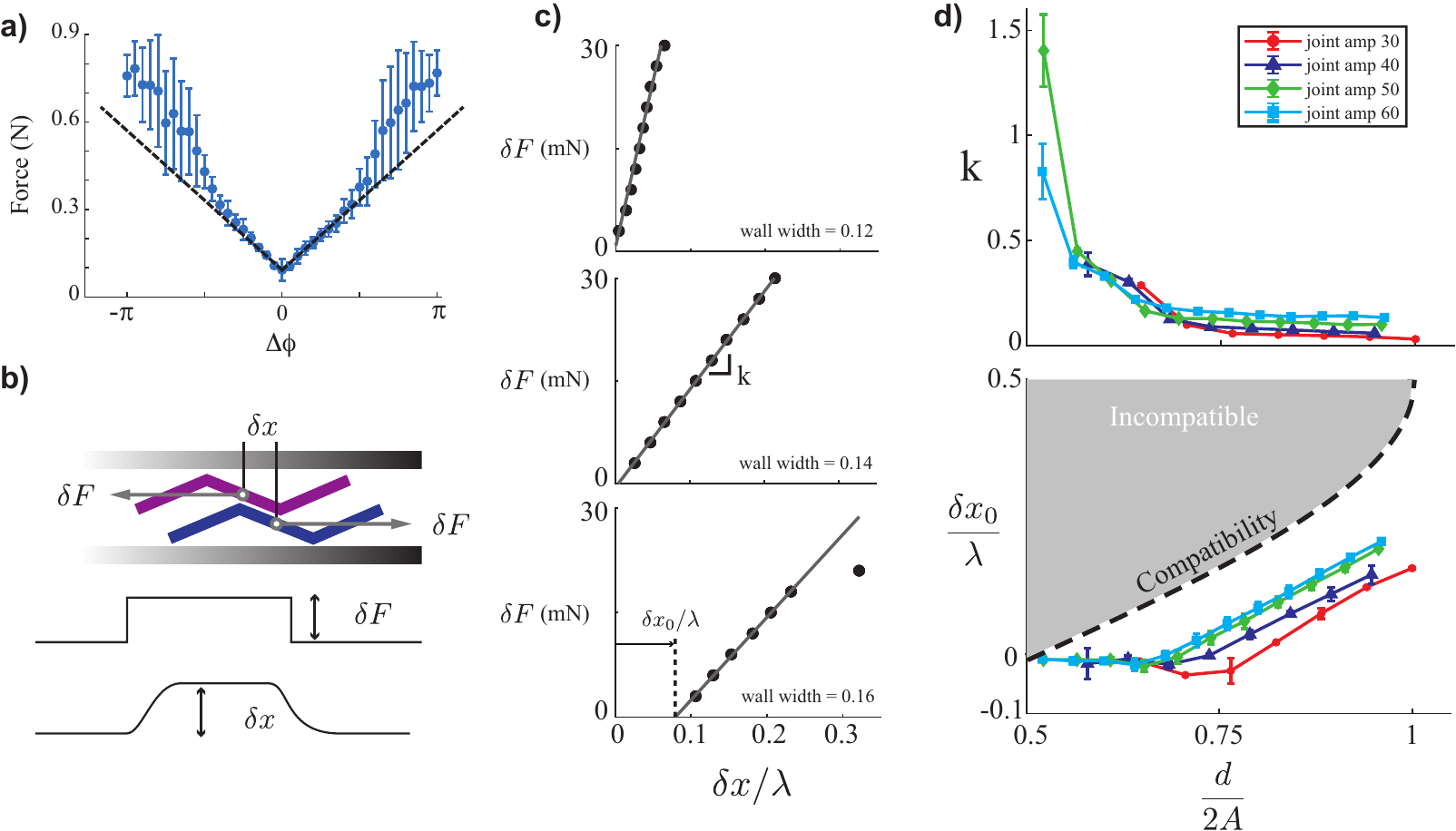}
    \caption{Cohesive longitudinal interactions depend on confinement. a) Magnitude of contact force between two swimmers during first two periods of oscillation, with initial conditions $\Delta x = 0$ and $\Delta \phi$ varied from $-\pi$ to $\pi$. The contact interaction force grows linearly with compatibility detuning. b) The cohesive magnitude of the compatible configuration was measured by applying equal and opposite perturbation forces, $\delta F$. The separation distance from compatibility, $\delta x$, is measured. c) Force-displacement relationship for three confinement distances. d) Effective interaction spring constant and offset distance ($\delta x_0$) as a function of confinement. Legend indicates different amplitude actuation.} 
    \label{fig:potential}
\end{figure*}

To observe how the compatibility states change when the swimmer separation distance is increased we performed simulations for $\Delta \phi = 0$ with initial conditions ranging from $\frac{\Delta x}{\lambda} \in [-1, 1]$.
We observed the final longitudinal separation distance as a function of the initial conditions and confinement. 
We normalize the confinement wall distance, $d$, by the peak-to-peak oscillatory amplitude of the undulatory body wave such that when $\frac{d}{2A} > 1$ swimmers through initial contact can push each other away and out of compatibility without the wall confinement to bring them back into contact.
In theory, confinement distances $\frac{d}{2A} < 0.5$ are not possible because the undulatory gait is obstructed however in practice because of the finite thickness of the swimmer cross section we restricted $\frac{d}{2A} \in [0.6, 1.2]$.
Increasing the confinement wall distance increased the range of final longitudinal spacing observed between two swimmers (Fig.~\ref{fig:spacing_basins}).
The central compatibility state broadened indicating that swimmers could compatibly move within a range of longitudinal separations without contact.
This is in accord with the predictions from the compatibility model (Eqn.~\ref{eqn:fullcompatibility}) in which non-zero lateral separations ($\Delta y > 0$) allow for a range of solutions to the compatibility conditions. 
In practice there will always be lateral spacing in active assemblies and this highlights a novel aspect of contact-mediated interactions because once in the compatible state the swimmers no longer can interact until pushed back out of compatibility.

\subsection{Potential energy modeling of compatible configurations}
\label{sec:potential}

In this last section we study the cohesive effect of gait compatible on the longitudinal dynamics between active undulatory systems. 
We first observe that the time-averaged contact forces between two swimmers out of compatibility.
The cumulative contact forces over the first two periods of oscillation are linearly dependent on the detuning phase difference, $\Delta \phi$, between the two swimmers (Fig.~\ref{fig:potential}a).
Dashed lines in Fig.~\ref{fig:potential}a are drawn to guide the eye and are symmetric about the positive and negative $\Delta \phi$. 

We hypothesize that the time-averaged contact dynamics between swimmers can be considered as an effective potential energy driving the pair to compatibility. 
To measure the effective potential of the compatibility configurations we allowed two swimmers with $\Delta \phi = 0$ to reach compatibility and then we applied a constant longitudinal separating force, $\delta F$, to each swimmer in opposing directions (Fig.~\ref{fig:potential}b). 
We observed that the swimmers separated by a longitudinal distance $\delta x$ in the presence of this force and for small $\delta F$ this position was sustained until the force was removed (Fig.~\ref{fig:potential}b). 
Thus, through the force perturbation we can probe the potential energy basin of the compatible configuration for $\Delta \phi = 0$.

We varied $\delta F$ and measured $\delta x$ over a range of confinement wall distances to observe the cohesive interaction. 
We observed a linear relationship between the applied force and the steady state separation suggesting that compatible configurations act like a simple harmonic potential (Fig.~\ref{fig:potential}c). 
We fit the ``stiffness'' of the compatible state as $\delta F = k (\delta x - \delta x_0)$ where $k$ is the interaction potential and $\delta x_0$ is an offset. 
The offset $\delta x_0$ represents the ability for systems with large enough lateral spacing to be found over a range of longitudinal separation distances ($\Delta x$) when in compatibility.
We performed this measurement over different swimming amplitudes and confinement distance. 

We found that the cohesion stiffness of the compatible state decreases with increasing wall width and becomes increasingly large as the confining wall spacing decreases (Fig.~\ref{fig:potential}d). 
Furthermore, the range of displacements with zero interaction force ($\delta x_0 > 0$) qualitatively follows the compatibility condition (Eqn.~\ref{eqn:fullcompatibility}) which is shown in the dashed line of Figure~\ref{fig:potential}d. 
These experiments indicate that compatible states are neutrally stable configurations and a linear interaction force drives swimmers into compatibility. 
The neutral region, $\delta x_0$, likely lies below the prediction from the compatibility model due to the three-link geometry of the swimmers.
This is similar to the comparison of the sinusoidal and three-link compatibility basins in which effects of the discrete link geometry cause the observations to deviate from theory.

We performed these simulations over a range of oscillation amplitudes. 
Increasing amplitude caused larger contact force interactions between swimmers and a larger projection of interaction surface along the longitudinal direction. 
Thus, larger amplitudes increase the stiffness of separation for a fixed wall width. 
However, when we normalize the wall width by gait amplitude the stiffness curves follow a similar trend (Fig.~\ref{fig:potential}d).

\section{Summary and conclusions}

In this paper, we studied the role of contact interactions between undulatory swimmers in experiment and simulation. 
We found that contact interactions among confined swimmers drive them to stable spatial configurations called compatible gaits (originally introduced in \cite{Yuan2014-xs}).
The compatibility criteria is determined by the lateral spacing and phase difference $\Delta\phi$ between swimmers. 
We found considerable agreement between the compatibility model prediction and the experiment and simulation results. 
Compatible gaits are equilibrium configurations with time-averaged interactions that have a linear force-displacement relationship along the longitudinal axis and are approximated as a harmonic potential well. 

Similar cohesive interactions have been observed to occur in other active, collective systems. 
However, such interactions are often mediated through a fluid and thus the interaction forces on the bodies can be exerted over long distances and smoothly decay as separation increases.  
For example recent work has found that the spatial arrangements of undulatory swimmers inspired from fish schools are cohesive dependent on actuation dynamics and spatial positioning \cite{Heydari2020-nr}.
Linear perturbations of these simple swimmers yielded linear interaction forces that were approximated as harmonic potentials \cite{Heydari2020-nr}.
Experiments with tandem undulating foils also demonstrate stable spatial configurations of the foils mediated through fluid mechanics \cite{Newbolt2019-gg} and stable but discrete swimming speeds \cite{Becker2015-xt}.
A fundamental difference in contact-coupled systems is that once out of reach, interactions can no longer occur. 
In theory long-slender swimmers in gait compatibility could be infinitesimally close and yet not have any physical interactions because they do not make contact.

This work was inspired in part by the observations of small undulatory worms and their collective swimming behaviors when in close proximity. 
Pairs of the nematode \textit{C. elegans} were studied in a confined channel (much like in the experiments reported here) and were observed to adjust their undulatory gait to match their neighbors \cite{Yuan2014-xs}.
The authors in that work argue that the nematodes are too large for hydrodynamics to be important and thus it must be contact that is driving the gait dynamics. 
We acknowledge there are competing theories for why such synchronization occurs in these worms \cite{Chakrabortty2019-qd} however these counterpoints do not contradict the work we have presented here. 
Similar collective undulatory gaits have been observed in the vinegar eels (\textit{Tubatrix aceti}) \cite{Quillen2021-fb} in which case the authors present a modified Kuramoto model that emulates the effect of steric interactions to describe the gait adjustment. 
In this work we do not allow for phase modulation of the swimmers, and only spatial rearrangement. 
However, this scenario is quite similar to the studies of infinitely long oscillating sheets that do not change undulatory phase but can re-align spatially and which are commonly referred to as ``synchronized'' in the literature \cite{Elfring2009-ze}.

This work is inspired from the broad areas of active matter systems, granular materials, and robotics. 
The convergence of these themes have been of significant interest in recent years because the stochastic behaviors of interacting robotics systems can be exploited for robust, redundant, and resilient robots.
Recent studies of robotic active matter such as smarticles \cite{Savoie2019-gx} and stochastic particles \cite{Li2019-ge} has highlighted how emergent collective behaviors can be designed and tuned through local contact interaction rules.
The role of contact interactions among shape-changing active matter systems may have applications in designing collective robot swarms that operate in close proximity. 
Building large functional systems from many constituent parts is not new in robotics and has gone under the titles of modular, reconfigurable, and swarm robotics over the years (See \cite{Seo2019-jy} and \cite{Brambilla2013-yl} for reviews).
However, recent connections drawn between these robotic collectives and active matter physics \cite{Fruchart2021-gf, Chvykov2021-ll} suggest novel and fruitful intersections between these fields in the years to come.

\bibliographystyle{apsrev4-2}
% \bibliography{refs} 
%%%%%%%%%%% .BBL file copied 


\end{document}